# SIMULATION OF THE THERMOELECTRIC EFFECT IN A MULTI-METALLIC SUPERCONDUCTING CAVITY *

N. K. Raut†, N. Stilin, J. Vennekate, S. Wang, G. Ciovati
Jefferson Lab, Newport News, VA, USA

## Abstract

Superconducting radio-frequency accelerating cavities made with different material layers, such as copper, Nb or Nb$_3$Sn, are susceptible to thermoelectric effects due to differences in Seebeck coefficients between the metals. A temperature gradient across the surfaces can drive thermoelectric currents, which may impact the cavity performance. A layered Cu/Nb/Nb$_3$Sn single-cell cavity was tested with cryocoolers in 2022. Three heaters were mounted on the cavity surface at different locations and three single-axis cryogenic fluxgate magnetometers were attached close to the cavity equator. A linear increase in the magnetic field was measured while increasing the heaters' power. The cavity setup was analysed with COMSOL and the results showed a trend similar to that observed in the experiment. This contribution details the approach chosen for the simulation and some of the challenges encountered.

# INTRODUCTION

Superconducting radio-frequency (SRF) cavities with high quality factor are used in particle accelerators, quantum computing, and the study of fundamental physics [1–4]. Niobium is the primary material of choice for these SRF cavities. Niobium-tin (Nb$_3$Sn) coated cavities have demonstrated comparable performance at higher cryogenic temperatures [5]. Recently, conduction-cooled SRF cavities have gained attention for industrial applications due to their compact design and cost effectiveness [6].

SRF cavities composed of different material layers, such as copper, Nb, or Nb$_3$Sn, are prone to thermoelectric effects due to variations in Seebeck coefficients [7]. Temperature gradients across these layers can induce thermoelectric currents that may degrade cavity performance. Figure 1 illustrates the concept of thermoelectric current generation in a multi-layered structure. A temperature gradient across materials with different Seebeck coefficients generates a potential difference, which in turn drives a thermoelectric current [8]. The equivalent circuit model demonstrates how the total current flow depends on both the temperature gradient and the electrical resistance of each layer.

This study presents a COMSOL Multiphysics simulation of thermoelectric behaviour in a layered conduction-cooled SRF cavity. A detailed comparison between the simulation and experimental results is also provided.

* This material is based upon work supported by the U.S. Department of Energy, Office of Science, Office of Accelerator R&D and Production, Accelerator Stewardship and Accelerator Development Program, and the Office of Nuclear Physics under contract DE-AC05-06OR23177.
† raut@jlab.org

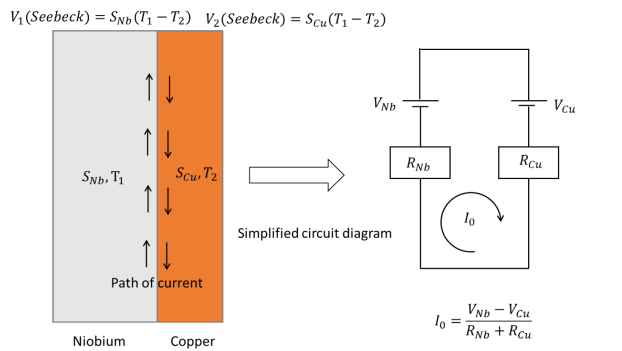

Figure 1: Conceptual illustration of thermoelectric current generation in layered materials with different Seebeck coefficients.

# 952.6 MHz MULTI-LAYER CAVITY

The experimental and simulated system implement a layered Cu/Nb/Nb$_3$Sn single-cell 952.6 MHz cavity [6]. Proceeding outward from the cavity interior, the inner surface consists of a ~3 $\mu$m thick Nb$_3$Sn layer, followed by a ~4 mm thick Nb layer and a ~7 mm thick Cu layer.

To enhance thermal conduction, a 19 mm thick OFHC copper ring was formed around the equator and a cooling plate of the same thickness was formed on one of the beam tubes. Commercial closed-cycle refrigerators known as cryocoolers are attached to the cooling plate and at two locations (top and bottom) of the cooling ring, through flexible Cu thermal straps. Figure 2 shows the positions of the heaters and of the three cryogenic single-axis fluxgate magnetometers (FGMs) used in the experiment, which were replicated in the simulation. Two heaters were mounted on the copper ring at the equator, and one on the beam tube section.

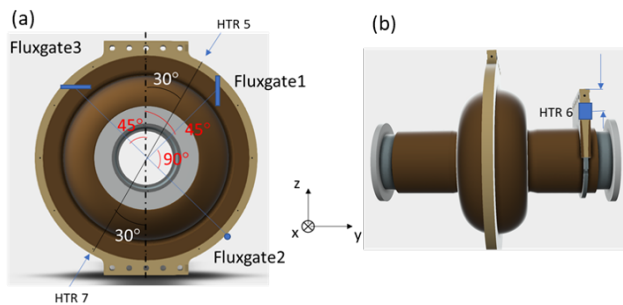

Figure 2: Layout showing the position of the heaters and FGMs used in the experiment (a)-(b).

## COMSOL MULTIPHYSICS

In the COMSOL Multiphysics simulation, heat sources were applied at the same surface locations as the experimental heaters to replicate the input power. The resulting temperature gradients generated thermoelectric currents, which were then coupled to magnetic field calculations.

Steady-state heat conduction in solids is modelled using:

$$\nabla \cdot (\kappa \nabla T) + Q = 0, \qquad (1)$$

where $\kappa$ is the thermal conductivity, $T$ is the temperature field, and $Q$ is the heat source.

The AC/DC module solves for the thermoelectric current density is computed as:

$$\mathbf{J} = -\sigma(\Delta V + S\,\Delta T). \qquad (2)$$

Here, $\sigma$ is the electrical conductivity, $S$ is the Seebeck coefficient, $V$ is the electric potential, and $T$ is the temperature.

This sequential multiphysics coupling captures the magnetic effects induced by thermally driven currents. Figure 3 shows the simulation flow: temperature gradients calculated in the heat transfer module are passed to the electric currents module, which computes the current density. This current is then used in the magnetic fields module to calculate the resulting field.

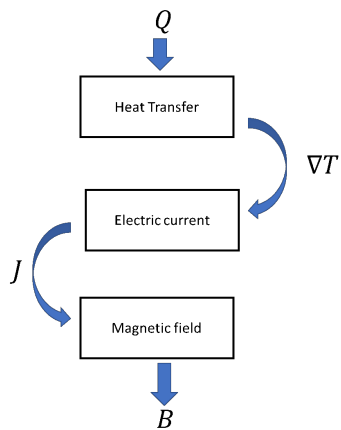

Figure 3: Simulation flow chart used for this work in COM-SOL.

Table 1: Material Property of Niobium and Copper

| Material | $\kappa[\mathrm{W/(m\cdot K)}]$ | $\sigma$ [S/m] | $S$ [V/K] |
|---|---|---|---|
| Nb | Fig. 3 | $1.0 \times 10^{12}$ | $1.0 \times 10^{-8}$ |
| Cu | 2230.0 | $6.0 \times 10^{7}$ | $1.6 \times 10^{-6}$ |

Table 1 summarizes the material properties used [7, 9]. A temperature-dependent thermal conductivity for niobium was applied, with experimental values shown in Fig. 4. Since the Nb is in the superconducting state, the electrical conductivity should be infinite and the Seebeck coefficient should be zero. In the simulation we have used a value of $\sigma$ much greater than that of Cu and a value of $S$ much smaller than that of Cu. The innermost $Nb_3Sn$ layer is omitted from the simulation, since its thickness and thermal conductivity are much smaller than that of the Nb layer, contributing less to the cavity temperature distribution.

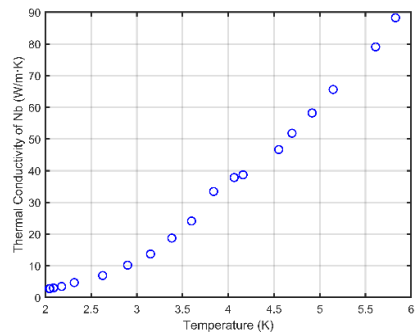

Figure 4: Measured thermal conductivity of niobium as a function of temperature.

In the simulation model, the copper tabs and heater-mounted regions are extruded to the faces of the bounding box and assigned as grounded domains as shown in Fig. 5(a). Each heater surface is defined as a terminal through which electrical power is supplied, while the three copper tabs, each held at 4.2 K, represent the cryocooler attachments, are set to ground potential. As shown in Fig. 5(b), the FGMs are modelled as lines in the simulation, each measuring the magnetic field along one axis. The position of the FGMs is known with an accuracy of 1 cm in the x-y plane. A global tetrahedral mesh is applied to the entire geometry, with element sizes ranging between 2 - 20 mm. To resolve the magnetic field around the fluxgate sensors, the mesh is locally refined: a work plane is placed through each fluxgate line and rotated by ±45°, creating three cut regions in the geometry. Those cut surfaces undergo two successive levels of uniform refinement. Finally, the OFHC copper surfaces at the equator receive one additional level of regular mesh refinement shown in Fig. 5(c).

## RESULTS AND DISCUSSION

Figure 6 compares simulated and measured magnetic field values at the three fluxgate locations for increasing heater power. The magnetic field for each FGM in the simulation is the average of the axial magnetic field along the line which represents the length of the FGM. The results from the simulations capture the linear trends observed experimentally, confirming that thermoelectric currents are the most likely cause for the measured changes in magnetic field with increasing heater power.

Discrepancies may arise from:
• Mesh resolution: Magnetic field calculations depend strongly on mesh density: coarse meshes produce fluctuating, noisy results, while extremely fine meshes push the simulation into a nonlinear region. Importantly, the trends remain fairly consistent between these two extremes, indicating convergence of the key physics.

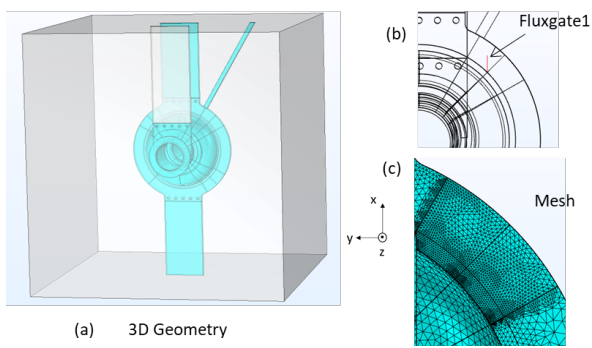

Figure 5: Experimental set up is replicated in simulation: (a) 3D geometry with bounding box where copper tabs and heater surfaces extend to the faces; (b) Representation of a straight-line fluxgate sensor in the model; (c) Locally refined mesh around each fluxgate.

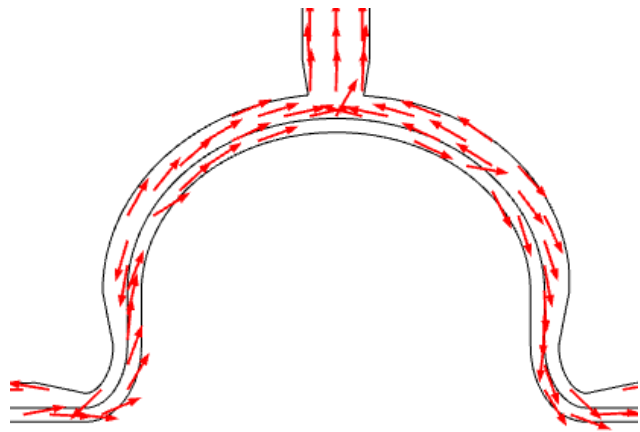

Figure 7: Result of COMSOL simulation showing thermoelectric current generation in the Nb-Cu layers near the cavity equator for heater 5 with 2 W of power.

• Uncertainty in the FGMs' positions: In the experiment, FGM mounting locations can vary by up to ≈1 cm between runs, and exact replication in the model is challenging. Even small location errors translate directly into field offsets.

• The presence of a residual magnetic field with unknown 3D distribution. The residual magnetic fields measured with no current applied to the heaters were 3.2 mG, 0.6 mG and 0.5 mG for FGM 1, 2 and 3 respectively. These residual mag-netic field values have been subtracted from the experimental values shown in Fig. 6.

Figure 7 displays, as an example, the thermoelectric current distribution in the Nb-Cu region for heater 5 with 2 W, showing opposing current paths in the two layers, in some regions of the model.

## CONCLUSION

This work presents a COMSOL Multiphysics simulation of thermoelectric effects in an Nb-Cu layered, conduction-cooled SRF cavity. The model qualitatively captures the linear increase of the magnetic field measured at three different locations and along three different axes, which was observed experimentally. Future studies may include a more complex simulation in which the RF power dissipated on the cavity inner surface is the heat source.

## ACKNOWLEDGEMENTS

We would like to thank A. Miyazaki of IJC Lab, France, for useful discussions on simulating thermoelectric currents in SRF cavities using COMSOL.

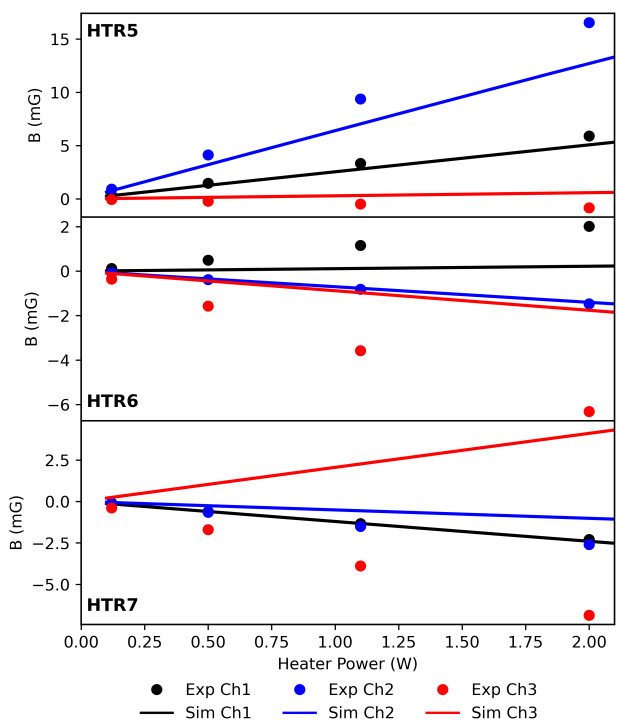

Figure 6: Comparison of experimental and simulated magnetic field measurements.